\documentclass[conference]{IEEEtran}
\IEEEoverridecommandlockouts
\usepackage{cite}
\usepackage{amsmath,amssymb,amsfonts}
\usepackage{algorithmic}
\usepackage{graphicx}
\usepackage{textcomp}
\usepackage{xcolor}
\def\BibTeX{{\rm B\kern-.05em{\sc i\kern-.025em b}\kern-.08em
    T\kern-.1667em\lower.7ex\hbox{E}\kern-.125emX}}

\usepackage{pgfplots}
\usepackage{tikz}
\usepackage{nicefrac}
\usepackage{adjustbox}
\usepackage{hyperref}
\usepackage{mathtools}

\DeclareMathOperator*{\minimize}{minimize}
\DeclareMathOperator*{\argmin}{arg\,min}
\DeclareMathOperator*{\argmax}{arg\,max}
\newcommand{\norm}[1]{\left\lVert#1\right\rVert}
\newcommand{\vect}[1]{\boldsymbol{\mathrm{#1}}}
\newcommand{\mat}[1]{\boldsymbol{\mathrm{#1}}}
\mathchardef\mhyphen="2D

\begin{document}

\title{Linear Precoder Design in Massive MIMO under Realistic Power Amplifier Consumption Constraint
\thanks{}
}

\author{\IEEEauthorblockN{Emanuele Peschiera}
\IEEEauthorblockA{\textit{ESAT-DRAMCO} \\
\textit{KU Leuven, Technology Campus Ghent}\\
9000 Ghent, Belgium \\
\href{mailto:emanuele.peschiera@kuleuven.be}{emanuele.peschiera@kuleuven.be}}
\and
\IEEEauthorblockN{François Rottenberg}
\IEEEauthorblockA{\textit{ESAT-DRAMCO} \\
\textit{KU Leuven, Technology Campus Ghent}\\
9000 Ghent, Belgium \\
\href{mailto:francois.rottenberg@kuleuven.be}{françois.rottenberg@kuleuven.be}}
}

\maketitle

\begin{abstract}
The energy consumption of wireless networks is a growing concern. In massive MIMO systems, which are being increasingly deployed as part of the 5G roll-out, the power amplifiers in the base stations have a large impact in terms of power demands. Most of the current massive MIMO precoders are designed to minimize the transmit power. However, the efficiency of the power amplifiers depend on their operating regime with respect to their saturation regime, and the consumed power proves to be non-linearly related to the transmit power. Power consumption-based equivalents of maximum ratio transmission, zero-forcing, and regularized zero-forcing precoders are therefore proposed. We show how the structure of the solutions radically changes. While all antennas should be active in order to minimize the transmit power, we find on the contrary that a smaller number of antennas should be activated if the objective is the power consumed by the power amplifiers.

\end{abstract}


\section{Introduction}
Information and communication technologies consumed approximately 1400 TWh of electricity in 2021, which represents the 6-7\% of the world electricity demand \cite{power-ICT-1}.
Within the mobile networks, base stations (BSs) are accountable for more than 50\% of the total consumption on average \cite{power-ICT-2}, even though this consumption depends on the application. Therefore, it is important to include realistic power consumption criteria into the design of communication systems.

The current state of the art linear massive MIMO precoders, e.g., maximum ratio transmission (MRT), zero-forcing (ZF) and regularized zero-forcing (RZF) \cite{mMIMO}, are obtained under a total transmit power constraint and not a total consumed power constraint. Indeed, these precoders minimize the total transmit power given that certain user requirements are satisfied. Implicitly, the consumed power $p_{\mathrm{cons}}$ is assumed to be linearly proportional to the transmit power $p_{\mathrm{tx}}$, i.e.,
\begin{equation}
p_{\mathrm{cons}} = p_{\mathrm{tx}}/\eta,
\end{equation}
$\eta$ being the power amplifier (PA) efficiency, which is assumed to be constant. However, it is known that the actual PA efficiency is not constant but depends on the ratio between the output power and the maximum output power, i.e., the back-off \cite{RF-PAs}. It is generally agreed that the PA causes most losses in the transmitter \cite{cft-rappaport}, therefore its realistic behavior needs to be taken into account. Given a BS equipped with $M$ antennas, and defining $p_m$ as the output power at the PA $m$, the total transmit power is $p_{\mathrm{tx}} = \sum_{m=0}^{M-1}p_m$. A more realistic model of the efficiency of the PA of antenna $m$ is given by \cite{p-cons-MIMO-1}
\begin{equation}\label{eta}
\eta_m = \eta_{\mathrm{max}}\sqrt{\frac{p_m}{p_{\mathrm{max}}}},
\end{equation}
where $p_{\mathrm{max}}$ is the maximal PA output power and $\eta_{\mathrm{max}}$ is the maximal PA efficiency, obtained for $p_m=p_{\mathrm{max}}$, which is, e.g., around 78.5\% for class B amplifiers. The model in \eqref{eta} has been shown to be accurate for many classes of amplifiers when the output power is
low \cite{RF-PAs}. The total consumed power becomes
\begin{equation}\label{CPM}
p_{\mathrm{cons}} = \sum_{m=0}^{M-1}p_m/\eta_m = \frac{\sqrt{p_{\mathrm{max}}}}{\eta_{\mathrm{max}}}\sum_{m=0}^{M-1}\sqrt{p_m}.
\end{equation}
Therefore, $p_{\mathrm{cons}}$ presents a square root dependence on $p_m$. As a result, common designs that minimize the total transmit power do not minimize the total consumed power. To the best of our knowledge, few works have taken this more advanced model into account for design precoders. The MISO case has been studied in \cite{p-cons-MIMO-1}. In \cite{p-cons-MIMO-2}, the problem was analyzed in detail from the capacity point of view in point-to-point MIMO, but only formulated for multi-user MIMO. No results and interpretations were proposed in the latter case.

In this work, we study novel precoders that minizime the consumed power subject to user performance criteria and a maximal per-antenna output power constraint. Our contribution can be summarized as follows:
\begin{itemize}
	\item We first show the expression of the consumed power in terms of the precoding coefficients. In the multi-user case, $p_{\mathrm{cons}}$ corresponds to the $L_{2,1}$ norm of the precoding matrix. In the single-user case, it simplifies to the $L_1$ norm of the precoding vector. These norms preserve convexity, which allows to obtain efficient solutions, and promote sparsity, which radically changes the structure of the solution compared to the conventional precoders.
	\item We then derive the power consumption-efficient equivalent of MRT precoder in the single-user scenario. The closed-form solution, which was previously obtained in \cite{p-cons-MIMO-1}, consequently saturates the antennas with the strongest channel gains until the required SNR is achieved. The gain in power consumption is evaluated varying the number of BS antennas.
	\item Last, we investigate the power consumption-efficient equivalents of ZF and RZF precoders in the multi-user scenario. In this case, convex optimization solvers allow to obtain the precoding coefficients. Compared to conventional precoders, a smaller number of antennas are activated. As the number of users grows, more antennas are utilized, and the gain in power consumption decreases.
\end{itemize}

\subsection{Notations}
Vectors and matrices are denoted by bold lowercase and uppercase letters, respectively. Superscripts $^*$, $^T$ and $^\dag$ stand for complex conjugate, transpose, and conjugate transpose operators. The symbols $\mathrm{tr}[\cdot]$ and $\mathbb{E}[\cdot]$ denote the trace and the expectation operators, while $\jmath$ is the imaginary unit. The notation 
$\mathrm{diag}(\vect{a})$ refers to a diagonal matrix whose $k$-th diagonal entry is equal to the $k$-th entry of the vector $\vect{a}$. A zero-mean complex Gaussian random variable with variance $\sigma^2$ is represented by $\mathcal{CN}\left(0,\sigma^2\right)$. A $K\times1$ all-zero vector is $\vect{0}_K$, while $\mat{I}_K$ is an identity matrix of size $K$. The $(i,j)$-th element of $\mat{A}$ is indicated by $[\vect{A}]_{i,j}$. Last, $\mat{A}^{1/2} = \mat{B} \iff \mat{A} = \mat{B}\mat{B}$, 
$\mat{A}^{n} = \underbrace{\mat{A}\mat{A}\dots\mat{A}}_{n \mathrm{\ times}}$, 
and $\mat{A}^{-n} = \underbrace{\mat{A}^{-1}\mat{A}^{-1}\dots\mat{A}^{-1}}_{n \mathrm{\ times}}$ for $n\in\mathbb{N}^+$.

\section{System Model}

\subsection{Signal Model}
We consider a massive MIMO system with $M$ antennas and $K$ users. The symbol of user $k$ is denoted by $s_k$, it has unitary power and different user symbols are assumed uncorrelated. The symbols are linearly precoded so that the transmitted signal $\vect{x} \in \mathbb{C}^{M\times1}$ is
\begin{equation}
\vect{x} = \mat{W}^T\vect{s},
\end{equation}
where $\mat{W} \in \mathbb{C}^{K\times M}$ is the precoding matrix and $\vect{s} \in \mathbb{C}^{K\times1}$ contains the transmitted symbols. 
The coefficient $\left[\vect{W}\right]_{k,m} = w_{k,m}$ represents the precoding weight for user $k$ at antenna $m$.
The received signal $\vect{r} \in \mathbb{C}^{K\times1}$ is
\begin{equation}
\vect{r} = \mat{H}\vect{x}+\vect{\nu},
\end{equation}
where $\mat{H} \in \mathbb{C}^{K\times M}$ is the channel matrix, $\left[\vect{H}\right]_{k,m} = h_{k,m}$ being the channel between user $k$ and antenna $m$, and $\vect{\nu} \sim \mathcal{CN}\left(\vect{0}_K,\sigma_\nu^2\mat{I}_K\right)$ represents the thermal noise.

\subsection{Consumed Power Model}
Being the symbols uncorrelated and with unit power, the transmit power at antenna $m$ is
\begin{equation}
p_m = \mathbb{E}\left[|x_m|^2\right] = \sum_{k=0}^{K-1}|w_{k,m}|^2,
\end{equation}
and the total transmit power becomes
\begin{equation}
p_{\text{tx}} = \sum_{m=0}^{M-1}\sum_{k=0}^{K-1}|w_{k,m}|^2 = \norm{\mat{W}}_F^2,
\end{equation}
where $\norm{\cdot}_F$ is the Frobenius norm. Using the model in \eqref{CPM}, the total consumed power by the PAs can be expressed as
\begin{equation}\label{L_21}
p_{\text{cons}} = \frac{\sqrt{p_{\mathrm{max}}}}{\eta_{\mathrm{max}}}\sum_{m=0}^{M-1}\sqrt{\sum_{k=0}^{K-1}|w_{k,m}|^2} = 
\frac{\sqrt{p_{\mathrm{max}}}}{\eta_{\mathrm{max}}}\norm{\mat{W}}_{2,1},
\end{equation}
where $\norm{\cdot}_{2,1}$ is the $L_{2,1}$ norm.

\subsection{Channel Model}
As channel models, we consider pure line-of-sight (LOS) and independent and identically distributed (i.i.d.) Rayleigh fading, which corresponds to non-line-of-sight (NLOS) scenarios\footnote{We did not include the large-scale fading as we make a first comparison between conventional and power consumption-efficient precoders, and we consider low output powers compared to the PA saturation power.}.
For LOS channels, and assuming a uniform linear array (ULA) with inter-antenna spacing of half a wavelength,
\begin{equation}
\left[\mat{H}\right]_{k,m} = e^{-\jmath \pi \cos{\theta_k}m },
\end{equation}
where $\theta_k$ is the user angle with respect to the array axis. In NLOS channels,
\begin{equation}
\left[\mat{H}\right]_{k,m} = h_{k,m},
\end{equation}
where $h_{k,m} \sim \mathcal{CN}\left(0,1\right)$ is the small-scale fading coefficient, which is assumed uncorrelated to the other coefficients in $\mat{H}$.

\section{Single-User Scenario}
When $K=1$, the precoding matrix becomes a vector $\vect{w} \in \mathbb{C}^{1\times M}$. Accordingly, the Frobenius norm simplifies to the squared $L_2$ norm, i.e., $p_{\mathrm{tx}} = \norm{\vect{w}}_2^2$, while the $L_{2,1}$ norm becomes the $L_1$ norm, i.e., $p_{\mathrm{cons}} = \norm{\vect{w}}_1$. In the following, the index $k$ is omitted for clarity. Since no inter-user interference is present, the main parameter is the signal-to-noise ratio (SNR), defined as
\begin{equation}
\mathrm{SNR} = \frac{\left|\sum_{m=0}^{M-1} h_mw_m\right|^2}{\sigma_\nu^2}.
\end{equation}
The conventional MRT can be found by minimizing the total transmit power under a received SNR constraint $\gamma$
\begin{equation}
\begin{split}
    \minimize_{\{w_m\}}  \;\;\;\;\;  				&			p_{\mathrm{tx}} = \norm{\vect{w}}_2^2 \\
    \mathrm{subject\ to} \;\;\;\;\;  				&	 		\frac{\left|\sum_{m=0}^{M-1} h_mw_m\right|^2}{\sigma_\nu^2} \geq \gamma.
\end{split}
\end{equation}
While the objective function is convex, the constraint is not convex. Anyway, we can consider $\sum_{m=0}^{M-1} h_mw_m$ to be purely real and positive as this would imply the multiplication by a constant phasor (done at the receiver to perform coherent combining) and would not change both the objective function and the constraint. The problem becomes
\begin{equation}
\begin{split}
    \minimize_{\{w_m\}}  \;\;\;\;\;  				&			p_{\mathrm{tx}} = \norm{\vect{w}}_2^2 \\
    \mathrm{subject\ to} \;\;\;\;\;  				&	 		\sum_{m=0}^{M-1} h_mw_m \geq \gamma^{1/2}\sigma_{\nu}, 
\end{split}
\end{equation}
which is a convex problem and can be solved, e.g., exploiting the Karush–Kuhn–Tucker (KKT) conditions. The well-known MRT precoder is thus obtained, i.e., 
$\vect{w}^{\mathrm{MRT}} = \sqrt{p_{\mathrm{tx}}}\,\vect{h}^*/\norm{\vect{h}}_2$, which achieves $\mathrm{SNR}^{\mathrm{MRT}} = p_{\mathrm{tx}}\norm{\vect{h}}_2^2/\sigma_\nu^2$, where $\vect{h} \in \mathbb{C}^{1\times M}$ is the channel vector.

The power consumption-efficient version of the previous precoder, together with a maximal per-antenna power constraint (i.e., the PA output saturation power constraint), is
\begin{equation}\label{MRT-eff}
\begin{split}
    \minimize_{\{w_m\}}  \;\;\;\;\;  				&			p_{\mathrm{cons}} = \frac{\sqrt{p_{\mathrm{max}}}}{\eta_{\mathrm{max}}}\norm{\vect{w}}_1 \\
    \mathrm{subject\ to} \;\;\;\;\;  				&	 		\sum_{m=0}^{M-1} h_mw_m \geq \gamma^{1/2}\sigma_{\nu}, \\
    													&			|w_m|^2 \leq p_{\mathrm{max}} \; \forall m.
\end{split}
\end{equation}
The problem is convex, and therefore can be efficiently solved numerically \cite{cvxpy-1,cvxpy-2}. Moreover, a closed-form solution can be found through the KKT conditions by considering real non-negative precoding coefficients. Indeed, when the optimization variable is $\{g_m = w_me^{\jmath\angle{h_m}} \in \mathbb{R}$, $g_m \geq 0\}$, \eqref{MRT-eff} becomes differentiable. The optimal solution corresponds to pouring power in the antenna with strongest channel gain $|h_m|$ until saturation, then the second one, and so on until the target SNR is achieved\footnote{Note that the problem can be unfeasible if the target SNR is not achieved by setting all antennas to saturation.}. A similar result has been obtained in \cite{p-cons-MIMO-1}. Defining $\mathcal{K}_-$ and $\mathcal{K}_+$ as the sets of the antennas with the $L-1$ and $L$ strongest channel gains, we denote $\tilde{m} = \argmin_{m\in\mathcal{K}_+}|h_m|$. The precoding coefficients are then given by
\begin{equation}
w_m^{\mathrm{MRT\mhyphen eff.}} = e^{-\jmath\angle h_m}
\begin{cases}
      \sqrt{p_{\mathrm{max}}} 													& 		\; \mathrm{if} \; m \in \mathcal{K}_- \\
      \frac{\gamma^{1/2}\sigma_{\nu}-\zeta}{|h_{\tilde{m}}|} 				& 		\; \mathrm{if} \; m = \tilde{m} \\
      0 																				& 		\; \mathrm{otherwise},
\end{cases}
\end{equation}
where $\zeta = \sqrt{p_{\mathrm{max}}}\sum_{m'\in\mathcal{K}_-}|h_{m'}|$. Therefore, the antennas associated with the $M-L$ lowest channel gains remain inactive. This represents a major difference with respect to the conventional MRT, in which all the antennas are active provided that their channel gain is greater than zero. The transmit power required by the MRT is $p_{\mathrm{tx}}^{\mathrm{MRT}} = \gamma\sigma_\nu^2/\norm{\vect{h}}_2^2$, hence, using \eqref{L_21},
\begin{equation}
p_{\mathrm{cons}}^{\mathrm{MRT}} = \frac{\sqrt{p_{\mathrm{max}}}}{\eta_{\mathrm{max}}}\gamma^{1/2}\sigma_\nu\frac{\norm{\vect{h}}_1}{\norm{\vect{h}}_2^2}.
\end{equation}
In the power consumption-efficient MRT, when no maximal per-antenna power constraint is considered, all the power is allocated to antenna $\hat{m} = \argmax_{m}|h_m|$. The transmit power becomes $p_{\mathrm{tx}}^{\mathrm{MRT\mhyphen eff.}} = \gamma\sigma_\nu^2/|h_{\hat{m}}|^2$, and the consumed power is
\begin{equation}
p_{\mathrm{cons}}^{\mathrm{MRT\mhyphen eff.}} = \frac{\sqrt{p_{\mathrm{max}}}}{\eta_{\mathrm{max}}}\gamma^{1/2}\sigma_\nu\frac{1}{\norm{\vect{h}}_\infty}.
\end{equation}
where $\norm{\vect{h}}_\infty=|h_{\hat{m}}|$ is the $\infty$-norm of $\vect{h}$.
The power consumption gain (PCG) is therefore
\begin{equation}\label{PCG_SU}
\mathrm{PCG} = \frac{p_{\mathrm{cons}}^{\mathrm{MRT}}}{p_{\mathrm{cons}}^{\mathrm{MRT\mhyphen eff.}}} =
\norm{\vect{h}}_\infty\frac{\norm{\vect{h}}_1}{\norm{\vect{h}}_2^2}.
\end{equation}
For a pure LOS channel, $|h_m|=1 \; \forall m$, hence $\mathrm{PCG}=1$. If $\gamma$ is achieved, any power allocation gives the same result (the first constraint in \eqref{MRT-eff} reduces to $\sum_{m=0}^{M-1}g_m = \gamma^{1/2}\sigma_\nu$, which directly fixes the value of the cost function). When $|h_m|$ varies among the antennas, there is a gain in $p_{\mathrm{cons}}$.

\section{Multi-User Scenario}
Let us define the SINR of user $k$ as
\begin{equation}
\mathrm{SINR}_k = \frac{\left|\sum_{m=0}^{M-1}h_{k,m}w_{k,m}\right|^2}{\sum_{k'=0,k'\neq k}^{K-1}\left|\sum_{m=0}^{M-1}h_{k,m}w_{k',m}\right|^2+\sigma_\nu^2}.
\end{equation}
Given a target SINR $\gamma_k$ for each user, the conventional precoders can be found as a solution to
\begin{equation}
\begin{split}
    \minimize_{\{w_{k,m}\}}  \;\;\;\;\;  			&			p_{\mathrm{tx}} = \norm{\mat{W}}_F^2 \\
    \mathrm{subject\ to} \;\;\;\;\;  				&	 		\mathrm{SINR}_k \geq \gamma_k \; \forall k.
\end{split}
\end{equation}
The SINR constraints are not convex, but can be reformulated as convex \cite{tx-BF}, and the problem can be solved numerically.
The power consumption-efficient precoders, together with a maximal per-antenna power constraint, are designed by solving
\begin{equation}
\begin{split}
    \minimize_{\{w_{k,m}\}}  \;\;\;\;\;  			&			p_{\mathrm{cons}} = \frac{\sqrt{p_{\mathrm{max}}}}{\eta_{\mathrm{max}}}\norm{\mat{W}}_{2,1} \\
    \mathrm{subject\ to} \;\;\;\;\;  				&	 		\mathrm{SINR}_k \geq \gamma_k \; \forall k, \\
    													& 			\sum_{k=0}^{K-1}|w_{k,m}|^2 \leq p_{\mathrm{max}} \; \forall m.
\end{split}
\end{equation}
It is known how the $L_{2,1}$ norm promotes sparsity. Therefore, similarly to what happens in the single-user scenario, one can expect only a subset of BS antennas will be activated. However, it is important to remark how, to be able to spatially multiplex the $K$ users, at least $K$ antennas should be active. More in general, a decrease in the PCG is expected when the ratio $M/K$ decreases.

\subsection{Zero-Forcing}
The ZF precoder is found by imposing zero inter-user interference. Defining $\mat{D}_\gamma = \mathrm{diag}\left(\gamma_0,\dots,\gamma_{K-1}\right)$ as the diagonal matrix contaning the required SNRs of the users, the conventional ZF is found by solving
\begin{equation}
\begin{split}
    \minimize_{\{w_{k,m}\}}  \;\;\;\;\;  			&			p_{\mathrm{tx}} = \norm{\mat{W}}_F^2 \\
    \mathrm{subject\ to} \;\;\;\;\;  				&	 		\mat{H}\mat{W}^T = \mat{D}_\gamma^{1/2}\sigma_\nu.
\end{split}
\end{equation}
The above problem is convex and can be solved, e.g., using the Lagrange multipliers method, obtaining
\begin{equation}
\left(\mat{W}^{\mathrm{ZF}}\right)^T = \mat{H}^\dag\left(\mat{H}\mat{H}^\dag\right)^{-1}\mat{D}_\gamma^{1/2}\sigma_\nu.
\end{equation}
The power consumption-efficient ZF precoder, whose solution is denoted by $\mat{W}^{\mathrm{ZF\mhyphen eff.}}$, corresponds to
\begin{equation}\label{ZF-eff}
\begin{split}
    \minimize_{\{w_{k,m}\}}  \;\;\;\;\;  			&			p_{\mathrm{cons}} = \frac{\sqrt{p_{\mathrm{max}}}}{\eta_{\mathrm{max}}}\norm{\mat{W}}_{2,1} \\
    \mathrm{subject\ to} \;\;\;\;\;  				&	 		\mat{H}\mat{W}^T = \mat{D}_\gamma^{1/2}\sigma_\nu, \\
    													& 			\sum_{k=0}^{K-1}|w_{k,m}|^2 \leq p_{\mathrm{max}} \; \forall m.
\end{split}
\end{equation}
This problem also has a convex formulation. However, it is difficult to find a closed form solution, but it can be easily solved numerically.
To compare the two precoders, the PCG is computed as
\begin{equation}\label{PCG_ZF}
\mathrm{PCG} = \nicefrac{\norm{\mat{W}^{\mathrm{ZF}}}_{2,1}}{\norm{\mat{W}^{\mathrm{ZF\mhyphen eff.}}}_{2,1}}.
\end{equation}

\subsection{Regularized Zero-Forcing}
The conventional RZF precoder design problem is\footnote{We fixed the value of $\xi$ by using the KKT conditions for problem \eqref{RZF} such that to obtain the precoding matrix in \eqref{RZF_sol}. The proof is omitted for space reasons.}
\begin{equation}\label{RZF}
\begin{split}
    \minimize_{\{w_{k,m}\}}  \;\;\;\;\;  			&			p_{\mathrm{tx}} = \norm{\mat{W}}_F^2 \\
    \mathrm{subject\ to} \;\;\;\;\;  				&	 		\norm{\mat{H}\mat{W}^T-\mat{D}_\gamma^{1/2}\sigma_\nu}_F^2 \leq \xi,
\end{split}
\end{equation}
where $\xi = \sigma_\nu^2\,\mathrm{tr}\!\left[\mat{D}_\gamma\left(\frac{1}{\sigma_\nu^2}\mat{\Lambda}+\mat{I}_K\right)^{-2}\right]$ and the square matrix $\mat{\Lambda} = \mathrm{diag}\left(\lambda_0,\dots,\lambda_{K-1}\right)$ contains the eigenvalues of $\mat{H}\mat{H}^\dag$. Using the matrix eigendecomposition, $\mat{\Lambda}$ can be computed by solving $\mat{H}\mat{H}^\dag = \mat{U}\mat{\Lambda}\mat{U}^*$, $\mat{U}$ being the matrix having the eigenvectors of $\mat{H}\mat{H}^\dag$ on its columns.
The problem at hand is convex and its well-known solution is
\begin{equation}\label{RZF_sol}
\left(\mat{W}^{\mathrm{RZF}}\right)^T = \mat{H}^\dag\left(\mat{H}\mat{H}^\dag+\sigma_\nu^2\mat{I}_K\right)^{-1}\mat{D}_\gamma^{1/2}\sigma_\nu.
\end{equation}
The power consumption-efficient RZF is then given by
\begin{equation}
\begin{split}
    \minimize_{\{w_{k,m}\}}  \;\;\;\;\;  			&			p_{\mathrm{cons}} = \frac{\sqrt{p_{\mathrm{max}}}}{\eta_{\mathrm{max}}}\norm{\mat{W}}_{2,1} \\
    \mathrm{subject\ to} \;\;\;\;\;  				&	 		\norm{\mat{H}\mat{W}^T-\mat{D}_\gamma^{1/2}\sigma_\nu}_F^2 \leq \xi, \\
    													& 			\sum_{k=0}^{K-1}|w_{k,m}|^2 \leq p_{\mathrm{max}} \; \forall m.
\end{split}
\end{equation}
As for the ZF precoder, the problem is difficult to be solved analytically, despite having a convex formulation. However, numerical solvers can efficiently solve it. The obtained precoder can be compared to the one in \eqref{RZF_sol} using \eqref{PCG_ZF}.

\section{Numerical Results}
The numerical experiments have been carried out using as target SNR $\gamma_k = \gamma = 10$ dB, $\sigma_\nu=1$ and no maximal per-antenna power constraint. The constraint on $p_{\mathrm{max}}$ has been removed for comparison purposes, as standard precoders do not consider it. In the single-user scenario, only NLOS channels has been simulated, as in LOS channels eq.\ \eqref{PCG_SU} looses significance. Instead, in the multi-user scenario, both NLOS and LOS channels have been investigated.

\begin{figure}[htbp]
\centering
\begin{adjustbox}{width=.4\textwidth}
\includegraphics{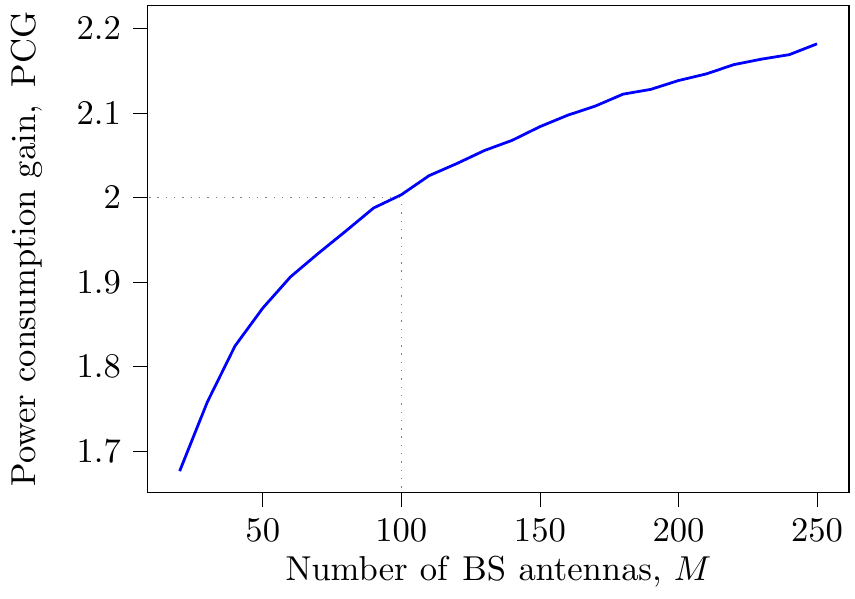}
\end{adjustbox}
\caption{Power consumption gain as a function of BS antennas for single-user case in NLOS channel, based on eq.\ \eqref{PCG_SU}. 
PCG was averaged over $10^{4}$ channel realizations.}
\label{fig1}
\end{figure}

In case of single-user systems, the results in terms of PCG are shown in Fig.\ \ref{fig1}. The PCG increases with $M$, as expected. Indeed, having no constraint on $p_{\mathrm{max}}$, only one antenna will be active in the power consumption-efficient precoder, regardless of $M$. Note that having a large $M$ is still important to increase spatial diversity in the selection of the antenna with the best channel gain. In the conventional MRT, instead, every additional antenna will be used, and therefore the gain in $p_{\mathrm{cons}}$ will be greater for larger numbers of BS antennas. For $M=100$ the PCG equals 2. Considering the constraint on $p_{\mathrm{max}}$, the PCG is expected to decrease when $p_{\mathrm{max}}$ is comparable to $\gamma$. More antennas will be saturated to achieve the SNR requirement, and this will reduce the gap between the two precoders.

\begin{figure}[htbp]
\centering
\begin{adjustbox}{width=.4\textwidth}
\includegraphics{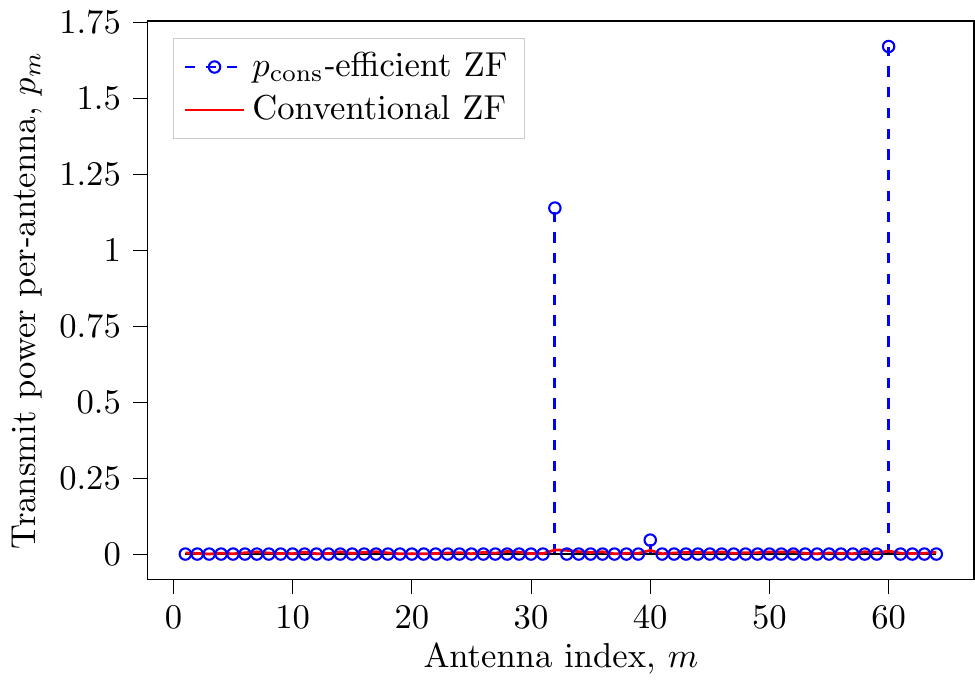}
\end{adjustbox}
\caption{Transmit power per-antenna for multi-user case with $K=2$ in NLOS channel. Conventional ZF (in red) and power consumption-efficient ZF (in blue) are shown.}
\label{fig2}
\end{figure}

\begin{figure}[htbp]
\centering
\begin{adjustbox}{width=.4\textwidth}
\includegraphics{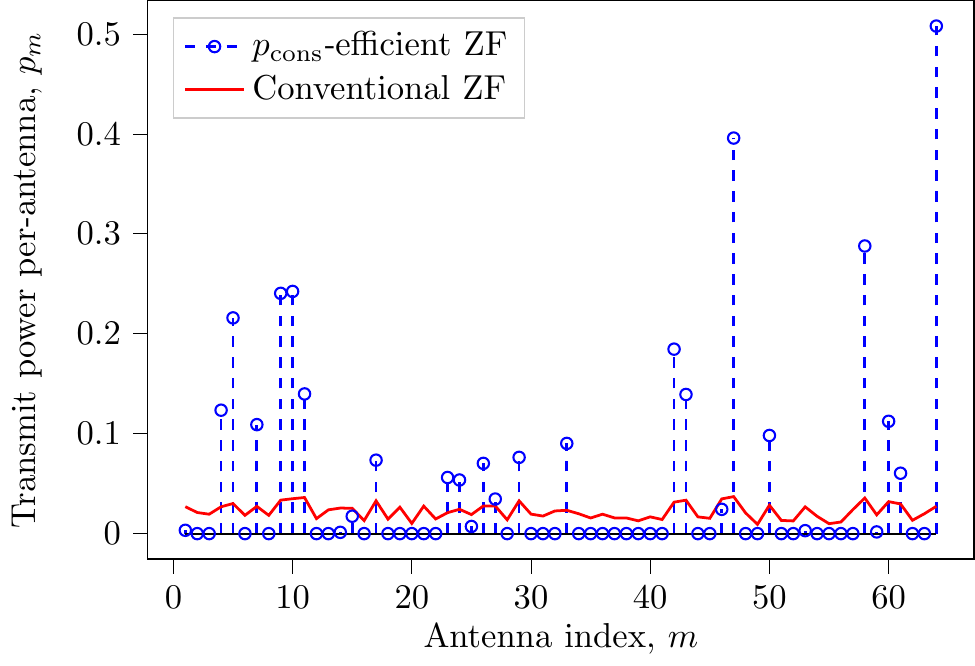}
\end{adjustbox}
\caption{Transmit power per-antenna for multi-user case with $K=8$ in NLOS channel. Conventional ZF (in red) and power consumption-efficient ZF (in blue) are shown.}
\label{fig3}
\end{figure}

\begin{figure}[htbp]
\centering
\begin{adjustbox}{width=.45\textwidth}
\includegraphics{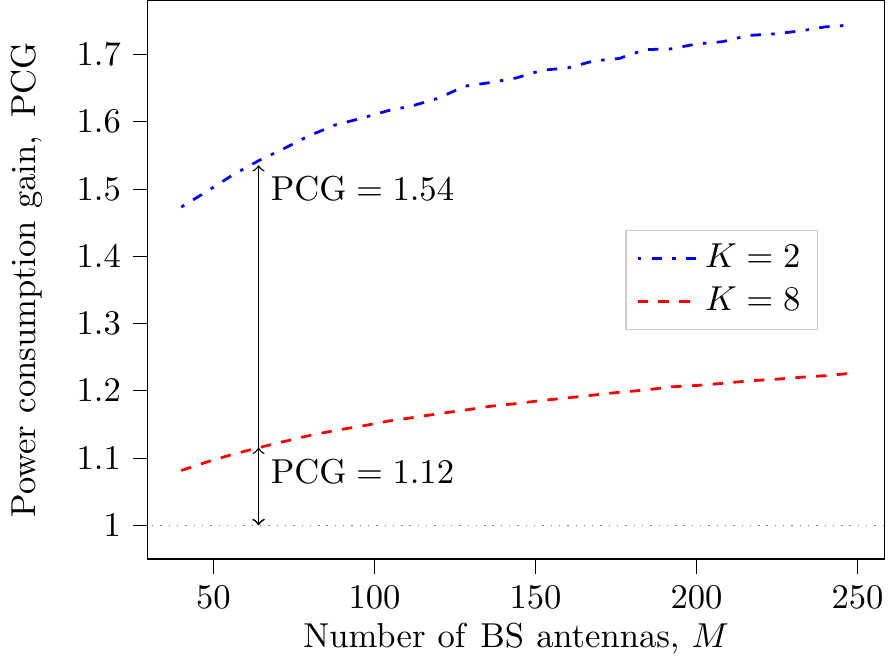}
\end{adjustbox}
\caption{Power consumption gain as a function of BS antennas for ZF precoder in NLOS channel, $K=2$ (in blue) and $K=8$ (in red). 
PCG was averaged over $10^{3}$ channel realizations.}
\label{fig4}
\end{figure}

For multi-user systems, the power consumption-efficient precoders also select a subset of the $M$ antennas. Fig.\ \ref{fig2} shows how, for a ZF precoder with $M=64$ and two users in a NLOS channel, only few antennas will have a non-zero power. When eight users are present (Fig.\ \ref{fig3}), more antennas will be activated in order to meet the SINR requirements of the users and mitigate inter-user interference. Nonetheless, the power distribution among the antennas varies significantly with respect to the conventional ZF scheme. About half of the antennas have $p_m \neq 0$ in the $p_{\mathrm{cons}}$-efficient ZF, while all the antennas are utilized in the conventional ZF. Figure \ref{fig4} illustrates the PCG as a function of $M$ in NLOS channels. For $K=2$, the PCG is still significant, e.g., it is equal to 1.54 when $M=64$. For eight users, instead, the PCG decreases and, when $M =64$, is equal to 1.12. Differently from the single-user scenario, the introduction of the constraint on $p_{\mathrm{max}}$ is here expected to have less impact (especially for large values of $K$). Indeed, multiple antennas have to be used in order to multiplex different users, thus their power will less likely be set closer to saturation.
In general, in LOS channels the PCG values are lower and close to one, i.e., no gain in power consumption. Similarly to the the single-user case, in which the $L_1$ norm and the squared $L_2$ norm are equivalent for unitary channel gains, the optimization on the $L_{2,1}$ norm and on the squared Frobenius norm gives similar results.

\section{Conclusion}
In this work, we conducted a first study on power consumption-efficient equivalents of the common massive MIMO linear precoders. A realistic expression of the power consumption, which is proportional to the square root of the transmit power, is first presented. The precoder design problems preserve convexity and can be easily solved numerically. In NLOS channels for both single-user and multi-user systems, the proposed precoders tend to activate few antennas compared to the conventional precoders. In LOS channels, given the same channel gains across both the antennas and the users, there are no significant differences. 
For the single-user scenario, a closed-form expression of the $p_{\mathrm{cons}}$-efficient precoder is illustrated, and the gain in power consumption is equal to 1.92 for $M=64$. Moving to the multi-user systems, the precoders design problem are first formulated and then solved numerically for a fixed SNR. For $M=64$, the gain in power consumption varies from 1.54 when $K=2$ to 1.12 when $K=8$. The gap between the $p_{\mathrm{cons}}$-efficient precoders and the transmit power-based precoders is not significant when several users are present. However, the structure of the retrieved solutions (e.g., Fig.\ \ref{fig3}) can provide further insights. Indeed, the entire RF chain of the non-active antennas can be turned off, and this will likely be associated with a higher PCG. More advanced models, e.g., including the circuit power consumption of each antenna, should be employed to have a complete characterization of the problem \cite{p-cons-Bjornson}. Future works also include numerical experiments considering the maximal per-antenna power constraint for both conventional and power consumption-efficient precoders. Non-linear distorsion terms caused by the PA can also be included in the precoder design, similarly to \cite{z3ro}. The analysis of multi-carrier systems would be of interest too, and cell-free deplyoments certainly need to be investigated.

\end{document}